\documentclass[floatfix,twocolumn,prd,showpacs,preprintnumbers,amsmath,nofootinbib,amssymb,noeprint]{revtex4-1}

\usepackage[utf8]{inputenc}
\usepackage{graphicx}
\usepackage{dsfont}

\usepackage{mathrsfs}
\usepackage{bm}
\usepackage{epstopdf}
\usepackage{color}
\usepackage{hyperref}
\hypersetup{colorlinks,citecolor=blue,urlcolor=blue,linkcolor=black}
\usepackage{braket}
\usepackage{placeins}
\usepackage{multirow}
\usepackage{float}
\usepackage{booktabs}
\usepackage{soul}
\usepackage{siunitx}
\usepackage{mathtools}
\usepackage{MnSymbol}
\usepackage[nameinlink,capitalize]{cleveref}

\def\gb{g_{\text{B}}}
\def\eb{e_{0}}

\def\Rep{\text{Re}}
\def\Imp{\text{Im}}

\def\tf{{\tau_{\text{F}}}}
\def\tauT{\tau T}

\def\CF{C_{F}}

\def\NT{N_{\tau}}

\newcommand{\diff}{\mathrm{d}}

\newcommand{\figurewidth}{0.425\textwidth}

\newcommand{\stackss}[2]{{\begin{subarray}{l}{\textrm{\tiny #1}}\\[-0.3ex] \textrm{\tiny #2}\end{subarray}}} 

\begin{document}

\title{Heavy quark momentum diffusion from the lattice using gradient flow}

\author{
   Luis Altenkort$^{\rm 1}$, Alexander M. Eller$^{\rm 2}$, O. Kaczmarek$^{\rm 1,3}$,
    Lukas Mazur$^{\rm 1}$, Guy D. Moore$^{\rm 2}$, H.-T. Shu$^{\rm 1}$
}

\affiliation{
    $^{\rm 1}$Fakult\"at f\"ur Physik, Universit\"at Bielefeld, D-33615 Bielefeld,
    Germany \\
    $^{\rm 2}$Institut f\"ur Kernphysik, Technische Universit\"at Darmstadt\\
Schlossgartenstra{\ss}e 2, D-64289 Darmstadt, Germany \\
    $^{\rm 3}$Key Laboratory of Quark and Lepton Physics (MOE) and Institute of
    Particle Physics, \\
    Central China Normal University, Wuhan 430079, China \\
}

\begin{abstract}
We apply the gradient flow on a color-electric two-point function that encodes the heavy quark momentum diffusion coefficient. 
The simulations are done on fine isotropic lattices in the quenched approximation at $1.5\,T_c$.
The continuum extrapolation is performed at fixed flow time followed by a second extrapolation to zero flow time. 
Perturbative calculations of this correlation function under Wilson flow are used to enhance the extrapolations of the nonperturbative lattice correlator. 
The final estimate for the continuum correlator at zero flow time largely agrees with one obtained from a previous study using the multilevel algorithm. 
We perform a spectral reconstruction based on perturbative model fits to estimate the heavy quark momentum diffusion coefficient. 
The approach we present here yields high-precision data for the correlator and is also applicable for actions with dynamical fermions. 
\end{abstract}


\maketitle
\section{Introduction}
The Yang-Mills gradient flow has proven to be a powerful tool for gauge theories since it was proposed in \cite{Narayanan:2006rf,Luscher:2009eq,Luscher:2010iy,Luscher:2011bx,Luscher:2013cpa}. 
Its applications on the lattice can be useful in several ways, for example reducing high frequency background noise from gauge configurations or setting the physical scale in lattice QCD \cite{Borsanyi:2012zs, Bazavov:2015yea, DallaBrida:2019wur}. 
Observables calculated under gradient flow are claimed to be automatically renormalized at sufficiently large flow time if the continuum limit is taken~\cite{Luscher:2010iy, Luscher:2010we}. 
This is a convenient property when dealing with operators whose renormalizations are troublesome, for instance the energy momentum tensor, which has already been studied under flow \cite{PhysRevD.94.114512,Kitazawa:2017qab, Suzuki:2013gza,Taniguchi:2016ofw}. Even more applications of the gradient flow can be found in \cite{Luscher:2013vga}.
 
Studies on transport coefficients like the heavy quark momentum diffusion coefficient, whose knowledge is of phenomenological interest, also benefit from the gradient flow. Nonperturbative determinations of such coefficients require high precision estimates of corresponding two-point correlation functions whose signals are often overshadowed by high frequency background noise even at large Monte Carlo sample sizes. Techniques like the multilevel algorithm~\cite{Luscher:2001up} and link-integration~\cite{DeForcrand:1985dr} can ameliorate the problem but are not applicable for actions with dynamical quarks. For now the gradient flow seems to be the only way to obtain high precision data of correlation functions in simulations with dynamical quarks. Discussions on how to flow dynamical quarks and their applications in dynamical QCD can be found in \cite{Luscher:2013cpa, Makino:2014taa, Taniguchi:2016ofw}.

In this paper we want to study how the well-known correlation function of color-electric fields which contains information about heavy quark transport behaves under Yang-Mills gradient flow and how to perform its continuum and flow-time-to-zero extrapolation. 
This color-electric correlator \cite{CaronHuot:2009uh} has been intensively studied in perturbation theory \cite{CasalderreySolana:2006rq, CaronHuot:2009uh, Eller:2018yje, Burnier:2010rp} and also nonperturbatively \cite{Meyer:2010tt, PhysRevD.85.014510, Francis:2015daa, Brambilla:2019oaa}. 
For this study we restrict ourselves to pure SU(3) gauge theory at a temperature of $T\approx 1.5\,T_c$. We compare our results with a previous nonperturbative measurement of the correlator from \cite{Francis:2015daa} that was obtained in the same setting but with the multilevel algorithm as a noise reduction method. This allows us to cross-check the results obtained from the gradient flow approach. 
The extrapolated correlator in the continuum limit at zero flow time is then used for spectral function reconstruction, for which we use multiple theoretically motivated fitting models, similar to what was done in \cite{Francis:2015daa}. 
Finally we compare the heavy quark momentum diffusion coefficient obtained from the extracted spectral function with those from other studies \cite{Francis:2015daa, Brambilla:2020siz}.

The paper is structured as follows: we start by recalling the definition of the gradient flow and how it is realized on the lattice (\autoref{sec:gradient-flow}). Afterwards we briefly present some perturbative calculations of the color-electric correlator at nonzero flow time and how we can use them to enhance the nonperturbative lattice data (\autoref{sec:EE-corr}). Section \ref{sec:EE-corr-nonpert} is then devoted to the analysis of this data and how we carry out the double extrapolation. The final estimate for the renormalized continuum correlator at zero flow time is then used to obtain the heavy quark momentum diffusion coefficient through spectral reconstruction in \autoref{sec:hq-mom-diff}.  We summarize our findings in \autoref{sec:conclusion}.

\section{Gradient Flow}\label{sec:gradient-flow}
The Yang-Mills gradient flow evolves the gauge field to some flow time $\tf$ along the gradient of the gauge action. 
The flowed gauge field $B_{\mu}(\tf,x)$ is defined through 
\begin{align}
    \label{boundary}
    B_{\mu}\Big{|}_{\tf=0}=A_{\mu}, \quad
    \frac{\partial{B_{\mu}}}{\partial{\tf}}=D_{\nu}G_{\nu\mu},
\end{align}
where $A_{\mu}$ is the ordinary gauge field. 
In order to fulfill gauge invariance one can easily construct the flowed covariant derivative and field strength following the standard Yang-Mills gauge theory:
\begin{equation}
    \begin{aligned}
        D_{\mu}&=\partial_{\mu}+[B_{\mu},{\makebox[1ex]{\textbf{$\cdot$}}}], \\ G_{\mu\nu}&=\partial_{\mu}B_{\nu}-\partial_{\nu}B_{\mu}+[B_{\mu},\ B_{\nu}].
    \end{aligned}
\end{equation}

In perturbation theory, a composite local operator $O(x,\tf)$ consisting of gauge fields can be expanded in $\tf$ as a superposition of the renormalized operator defined at zero flow time~\cite{Luscher:2011bx}:
\begin{equation}
    \label{short-time-expansion}
    O(x,\tf) \xrightarrow[\tf\rightarrow 0]{\text{}} \sum_i c_i(\tf)O^R_i(x).
\end{equation}
Here $c_i$ are some coefficients that can be calculated perturbatively. In order to obtain the renormalized operator in the original theory one needs to invert \cref{short-time-expansion} and go back to zero flow time. 
We will show how to carry out this procedure in practice in \autoref{sec:zero-flow-extr}.

To leading order a perturbative calculation in $D$ dimensions~\cite{Luscher:2010iy} provides a simple explanation for the effect of the flow on the gauge field:
\begin{align}
    \label{flow_solution}
        \begin{split}
        &B_{\mu}(x,\tf)=\int d^Dy\ K_{\tf}(x-y)A_{\mu}(y),\\
        &K_{\tf}(z)=\int \frac{d^Dp}{(2\pi)^D} e^{ipz}e^{-\tf p^2}=\frac{e^{-z^2/4\tf}}{(4\pi \tf)^{D/2}}.
    \end{split}
\end{align}
The flow averages the gauge field with a local Gaussian kernel. The mean-square radius of the Gaussian distribution is $\sqrt{8\tf}$ and is sometimes called the ``flow radius.''

How much flow can be applied to a gauge field before measuring a correlation function on the lattice? 
Consider the correlator of two operators whose temporal separation is $\tau$. 
On the one hand, the flow radius should be large enough to suppress  lattice discretization effects, and on the other hand, we need to make sure that the separation between operators is larger than the flow radius, so that the correlation between them is not contaminated. 
This leads to a general allowed flow time range of
\begin{align}
\label{flow_time_requirement}
a\lesssim \sqrt{8\tf} \lesssim \frac{\tau}{2}.
\end{align}
We can make the latter limit more rigorous by performing a perturbative determination of the flow radius where a specific correlator starts to suffer contamination due to the overlap of the operator smearing radii.  
We will do this for the color-electric correlator below.

\hspace{1cm}
\section{The color-electric correlator in perturbation theory}
\label{sec:EE-corr}

The correlator that we want to study under flow was first proposed in \cite{CaronHuot:2009uh} and is related to the thermalization rate of a heavy quark in a hot plasma. 
It arises from the infinite quark mass limit of the meson current-current correlator. 
The advantage of studying this specific correlator is that the spectral function encoded in it has no sharp transport peak, in contrast to the perturbative behavior for current-current and stress-stress correlators.
As a result, the relevant transport coefficient is much easier to obtain. 
The correlator is a product of electric fields that sit on a Polyakov loop, which is why it is referred to as the color-electric correlator. 
It is defined as \cite{CaronHuot:2009uh}
\begin{equation}
    G(\tau)=-\sum_{i=1}^3\frac{\langle {\rm{Re}\ \rm{Tr}}[U(\beta,\tau)E_i(\tau,\vec{0})U(\tau,0)E_i(0,\vec{0})]\rangle}{3\langle {\rm{Re}\ \rm{Tr}}[U(\beta,0)]\rangle},
    \label{gee}
\end{equation}
where $\beta=1/T$ is the temporal extent which is equal to the inverse temperature, and $U(\tau_2,\tau_1)$ is a Wilson line in the Euclidean time direction. 
$E_i$ is the color-electric field in the geometrical normalization, which we discretize following~\cite{CaronHuot:2009uh}:
\begin{equation}
    \label{ge}
    \begin{aligned}
        E_i(\tau,\vec{x})=&\ U_i(\tau,\vec{x})U_4(\tau,\vec{x}+\hat{i}) \\
        &-U_4(\tau,\vec{x})U_i(\tau,\vec{x}+\hat{4}).
    \end{aligned}
\end{equation}
Here, $U_i(\tau,\vec{x})$ is an SU(3) link variable in the $i$th direction.

The leading-order perturbative calculation in continuum of the correlator under gradient flow was performed in~\cite{Eller:2018yje}. 
We perform the analog calculation on the lattice here in order to illustrate the behavior of the correlator under Wilson flow compared to its continuum counterpart. 

\begin{widetext}
The LO perturbative lattice correlator under \mbox{Wilson} flow reads 
\begin{align}
    \label{glatflow1}
        & G^\stackss{norm}{latt}(\tau, \tf ) \equiv \frac{G^\stackss{LO}{latt}(\tau,\tf)}{g^2 C_F}= - \frac{1}{a^{3} \beta} \sum_{n=-\frac{\beta}{2}}^{\frac{\beta}{2}-1} \cos\left( 2\pi n \tau T \right)
        \\ \nonumber & \times
        \left[  e^{- 8\tf T^{2} \NT^{2}  \sin^{2}\left(\frac{\pi n}{\NT}  \right) } e^{- \frac{3}{2} 8\tf T^{2} \NT^{2}}  \left( I_{0}\left( {\scriptstyle \frac{ 8\tf T^{2} \NT^{2} }{2}}\right)\right)^{3} - \hspace{-0.5em} \int\limits_{8\tf T^{2} \NT^{2} }^{\infty} \hspace{-1em } \mathrm{d}x ~e^{- x  \sin^{2}\left(\frac{\pi n}{\NT}  \right) } e^{-\frac{3}{2} x} \left( I_{0}( {\scriptstyle\frac{x}{2}})\right)^{2} 
        \left\lbrace   I_{0}( {\scriptstyle \frac{x}{2}}) - I_{1}({\scriptstyle \frac{x}{2}}) \right\rbrace \phantom{\int\limits_{ }^{} \hspace{-1.0em}} \right]\,,
\end{align}
\end{widetext}
where $I_{0}\,,\,\,I_{1}$ are modified Bessel functions of the first kind
and $C_F= (N_c^2-1)/2N_c = 4/3$ is the quadratic Casimir of the fundamental representation.
Details of the derivation of this result can be found in the Appendix \ref{subsec:appendix-lo-derivation}.
We divide the perturbative correlator by $g^2 C_F$ and equip it with the superscript ``norm'' since we will later use it to normalize our nonperturbative lattice measurements.

For reference we also provide the leading-order lattice and continuum~\cite{CaronHuot:2009uh} color-electric correlator at \textit{zero flow time} here:
\begin{align}
&G^\stackss{norm}{latt}_{\tf=0}(\tau)=- \frac{1}{a^{3} \beta} \sum_{n=-\frac{\beta}{2}}^{\frac{\beta}{2}-1} \cos\left( 2\pi n \tau T \right)\\
&\times\left[1 - \hspace{-4pt} \int\limits_{0}^{\infty} \hspace{-0.5ex}\mathrm{d}x ~e^{- x  \sin^{2}\left(\frac{\pi n}{\NT}  \right) } e^{-\frac{3}{2} x}  \left( I_{0}({\scriptstyle \frac{x}{2}})\right)^{2} 
\left\lbrace   I_{0}({\scriptstyle \frac{x}{2}}) - I_{1}({\scriptstyle \frac{x}{2}}) \right\rbrace \phantom{\int\limits_{ }^{} \hspace{-1.0em}} \right],\nonumber
\end{align}
\begin{equation}
\begin{aligned}
    G^\stackss{norm}{cont}_{\tf=0}(\tau)&\equiv \frac{G^\stackss{LO}{cont}(\tau)}{g^2 C_F}=\hspace{-0.5pt}\pi^2T^4\hspace{-0.5pt}\left[\frac{\cos^2(\pi\tau T)}{\sin^4(\pi\tau T)} +\frac{1}{3\sin^2(\pi \tau T)}\right].\\
\end{aligned}
\end{equation}
\vspace{1ex}

\begin{figure}[!tb]
    \includegraphics[width=\figurewidth]{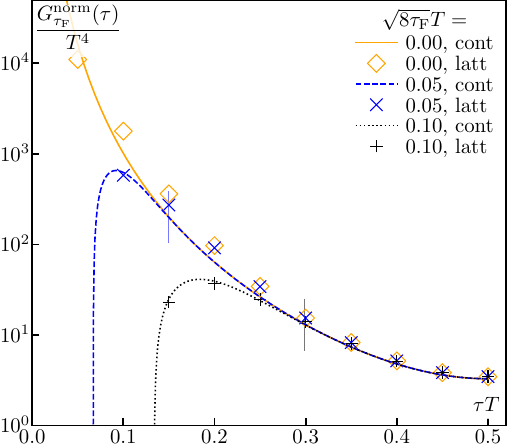}
    \caption{The color-electric correlator at different flow times calculated perturbatively to leading order in the continuum (``cont'', taken from \cite{Eller:2018yje}) and on the lattice (``latt'', \autoref{glatflow1} with $N_\tau = 20$).
    The vertical lines indicate the point where the flowed and nonflowed continuum results begin to agree to better than $1\%$.
    For the lattice correlators this distance needs to be shifted to the right by about one lattice spacing.}
    \label{fig:EE_flow_pert}
\end{figure}

In \autoref{fig:EE_flow_pert} we show an example for different flow times in the continuum and on a lattice with $N_{\tau}=20$. 
The vertical lines in this figure indicate the flow-time-dependent separation $\tau$ from which on the flowed continuum correlator deviates less than $1\%$ from its nonflowed counterpart; they are located at $\tau \approx 3\sqrt{8\tf}  $~\cite{Eller:2018yje}. By rearranging this relation we obtain the corresponding flow time for a given separation, which can serve as an upper limit.
For the lattice version of this limit we need to shift the allowed distance by about one lattice spacing, and together with \cref{flow_time_requirement} we obtain a serviceable flow interval of
\begin{align}
    \label{flow_time_full_limit}
    a \lesssim \sqrt{8\tf} \lesssim \frac{\tau-a}{3}.
\end{align}
Because of this limitation we can only hope to extract the large-distance behavior via the gradient flow method. 
Fortunately this is also the region that carries the information about the heavy quark momentum diffusion coefficient. 

From \autoref{fig:EE_flow_pert} we can see that the correlator grows as
$\tau^{-4}$ at small separations, and therefore spans several orders of
magnitude in value.
In order to improve visibility we can remove this dominant behavior from the nonperturbative lattice data that we present later by normalizing it with the perturbative leading-order correlator. 
The normalized data is also convenient for the extrapolations as its values are all of the same order of magnitude. 

Correlators calculated on the lattice suffer from cutoff effects. 
One can remove the leading-order contribution of these effects from the nonperturbative lattice data by utilizing the perturbative lattice and continuum correlators (see \cref{glatflow1} and \cite{Eller:2018yje}). 
This technique is known as tree-level improvement~\cite{Sommer:1993ce}. 
In Appendix \ref{sec:tree-level-imp} we explain in detail how to perform the improvement for the color-electric correlator under flow. 

Note that in our study the nonperturbative correlator is obtained using an improved discretization of the flow equation called \textit{Zeuthen flow} (see \autoref{sec:lattice-setup}), while the perturbative lattice calculations are done with \textit{Wilson flow}. 
The main point of appendix \ref{sec:tree-level-imp} is that it is reasonable to improve the Zeuthen-flowed correlator by using the nonflowed leading-order lattice correlator. 
Since we want to normalize the data with the leading-order continuum correlator, we end up with the dimensionless ratio [cf. \cref{tree-level-imp} and \cref{Gimp_zeuthen}]
\begin{align}
    \label{Gimp}
    \frac
    {G^\textrm{latt}_\tf(\tau)}
    {G_{\tf=0}^\stackss{norm}{latt}(\tau)},
\end{align}
on which the continuum and flow-time-to-zero extrapolations will be performed.

\section{Flowed color-electric correlator on the lattice}\label{sec:EE-corr-nonpert}

\begin{figure*}[!t]
    \centering
    \includegraphics[width=\figurewidth]{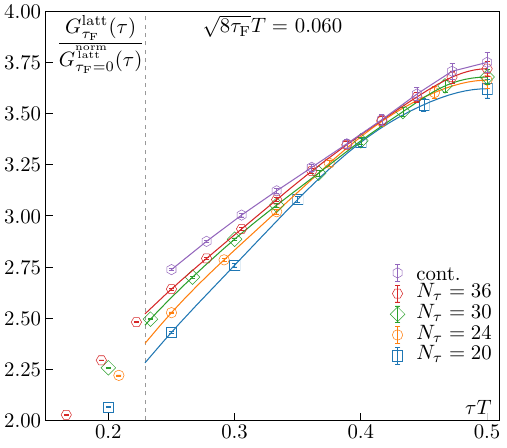}
    \hspace{1.5cm}
    \includegraphics[width=\figurewidth]{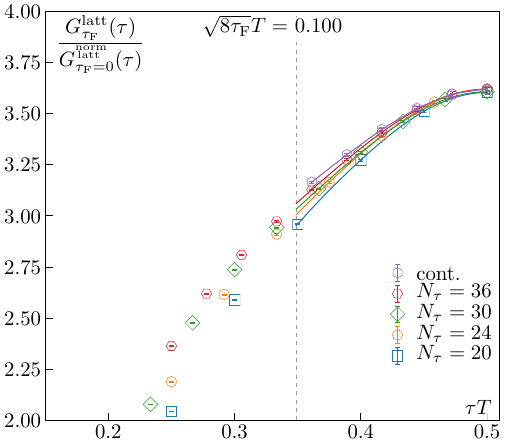}
    \caption{Left: the color-electric correlator at $T\approx 1.5\,T_c$ on different lattices at a small flow time ($\sqrt{8\tf}T=0.060$). To help guide the eye the interpolations going through the data points are shown and the continuum extrapolated points (see \autoref{sec:cont-extr}) are connected by straight lines. The dashed vertical line depicts the flow-time-dependent distance limit up to which we trust the lattice correlator (see \autoref{flow_time_full_limit}, here $a = 1/(N_\tau T)$ with $N_\tau = 20$). Right: the same but at $\sqrt{8\tf}T=0.10$.}
    \label{fig:EE-discretization-effects}
\end{figure*}

In this section we present our measurements and analysis of the color-electric correlator on fine isotropic lattices at $T\approx 1.5\,T_c$. We perform the continuum extrapolation at fixed flow time and subsequently the extrapolation to $\tf = 0$. 

\subsection{Lattice setup}\label{sec:lattice-setup}

\begin{table}[!b]
    \centering
    \begin{tabular}{ccrcccc}                            
    \hline \hline
    $a$ (fm) & $a^{-1}$ (GeV) & $N_{\sigma}$ & $N_{\tau}$ & $\beta$ & $T/T_{c}$ & \#conf.\tabularnewline
    \hline
    0.0262 & 7.534 & 64 &  16  & 6.8736 &  1.51  & 10000 \tabularnewline
    0.0215 & 9.187 & 80 &  20  & 7.0350 &  1.47  & 10000 \tabularnewline
    0.0178 & 11.11 & 96 &  24  & 7.1920 &  1.48  & 10000 \tabularnewline
    0.0140 & 14.14 & 120 & 30  & 7.3940 &  1.51  & 10000 \tabularnewline
    0.0117 & 16.88 & 144 & 36  & 7.5440 &  1.50  & 10000 \tabularnewline
    \hline \hline
    \end{tabular}
    \caption{Lattice spacings, lattice dimensions, $\beta$ values, temperature and number of configurations generated for this work. The lattice spacing $a$ is determined via the Sommer scale $r_0$~\cite{Sommer:1993ce} with parameters taken from~\cite{Francis:2015lha} and updated coefficients from~\cite{Burnier:2017bod}. We use $r_0T_c=0.7457(45)$~\cite{Francis:2015lha}. The coarsest lattice ($N_\tau=16$) does not enter in the continuum extrapolation. }
    \label{tab:lattice_setup}
\end{table}

A summary of the parameters of the gauge configurations used in this study can be found in \autoref{tab:lattice_setup}. 
All configurations are generated in the quenched approximation using heatbath and overrelaxation updates. 
The gauge action is the standard Wilson action. 
After thermalization (5000 heatbath sweeps), we save a configuration after every 500 combined sweeps, where one such sweep consists of one heatbath and four overrelaxation sweeps. 
We have checked that this procedure eliminates autocorrelations between saved configurations in all quantities we consider here.
The lattice boundary conditions are periodic for all directions. 

In order to compare our correlator measurements with previous results from \cite{Francis:2015daa}, we not only use the same gauge action and operator discretization, but also the same lattice dimensions and $\beta$-values to set the temperature to about $1.5\,T_c$. 
The lattice spacing $a$ is determined via the Sommer scale $r_0$~\cite{Sommer:1993ce} with parameters taken from~\cite{Francis:2015lha} and updated coefficients from~\cite{Burnier:2017bod}. 
We use the state-of-the-art value $r_0T_c=0.7457(45)$~\cite{Francis:2015lha}.

For statistical mean and error estimation of observables measured on the configurations we perform a bootstrap analysis with 10000 samples for each lattice.  

On the lattice we need to use the discretized counterparts of the continuum flow equations. 
Specifically, we use a Symanzik improved version of the flow called \textit{Zeuthen flow}~\cite{Ramos:2015baa}, which eliminates $\mathcal{O}(a^2)$ cutoff effects that are present in the ordinary Wilson flow. 
The Zeuthen flow also seems to be more stable at larger flow times.
We solve \autoref{boundary} numerically using an adaptive step-size Runge-Kutta method for Lie groups~\cite{CELLEDONI2003341}.
The observables are measured on flow times $\sqrt{8\tf}T\in \lbrace 0, 0.001, \dots, 0.199,0.2\rbrace$.

\subsection{Cutoff effects and signal-to-noise ratio under flow}
\label{sec:Cutoffeffects}

Calculations of the color-electric correlator without gradient flow exhibit huge statistical errors even for the large amounts of gauge configurations used in this study (see \autoref{tab:lattice_setup}). 
Beyond the first few lattice spacings the obtained data is almost pure noise at zero flow time. 
By increasing the flow time, high-frequency fluctuations of the gauge fields are suppressed and the signal-to-noise ratio of the correlator is expected to increase. 
Additionally, cutoff effects are expected to decrease with increasing flow time.

The minimum amount of flow that is needed to produce renormalized operators is $\sqrt{8\tf}T \approx aT = N_\tau^{-1}$. Via \cref{flow_time_full_limit} one can convert this minimum flow time into a lower bound for the separation $\tau T$. Because of this we henceforth exclude the $N_\tau=16$ lattice (see \autoref{tab:lattice_setup}) from the analysis. 
The limits are then dictated by the $N_\tau=20$ lattice, and so we have to flow up to at least $\sqrt{8\tf}T = 0.05$,
and we can only obtain renormalized correlator data for $\tau T \geq 0.2$.
These limits are also applied to the finer lattices and will carry over to the continuum and flow-time-to-zero extrapolations.

In \autoref{fig:EE-discretization-effects} we show the tree-level improved correlators on all lattices at one intermediate ($\sqrt{8 \tf} T = 0.06$) and one larger flow time ($\sqrt{8 \tf} T = 0.10$). 
The dashed vertical line depicts the lower boundary for the distance resulting from the upper boundary for the flow time [\cref{flow_time_full_limit}].
As expected, the flow drastically improves the signal and ameliorates cutoff effects, but it also contaminates the correlation at smaller distances if too much flow is applied. 
The continuum values shown in these figures will be explained in the following section.

\autoref{fig:EE_flow_effect} provides a more detailed look into how the correlator behaves under flow at fixed separations on the finest available lattice ($N_\tau =36$). 
After the initial ``signal acquisition phase'' there is only a minor effect that the flow exerts on the longer-distance correlation. 
The second lattice separation is particularly interesting since it is large enough to prevent immediate operator contamination but small enough to inherit almost no noise at zero flow time. 
At this separation we can see an initial rising behavior under flow, which we attribute to the renormalization caused by the flow. 
(This behavior can be observed on all lattices of \autoref{tab:lattice_setup}; the rising behavior is not visible for the higher separations as here the dominant effect of the flow is the improvement of the signal.)
A similar renormalization-induced behavior can also be observed for the electric correlator in perturbative lattice QED under Wilson flow, for which we explicitly calculate the ``tadpole-type'' next-to-leading order contributions in Appendix \ref{sec:appendix-renorm-flow}. 
Our understanding of the behavior in \autoref{fig:EE_flow_effect} is that after some flow, the renormalization is complete and the correlation then behaves almost linearly for some time, before it is then substantially contaminated.
After a continuum extrapolation we will exploit the seemingly linear region to extrapolate the correlator to zero flow time. 
The dominant flow-time-dependent contributions in this region will essentially consist of gradually accruing contact terms of the operator. 
In principle these exist as soon as the flow time is nonzero, and their contribution is enhanced by the fact that the operator is not truly local due to the discretization of the electric fields [\cref{ge}].

\subsection{Continuum extrapolation}\label{sec:cont-extr}

The physical correlation function is at vanishing lattice spacing and flow time, so we need to perform a double extrapolation. 
But it is important
to perform the extrapolations in the right order:  the continuum extrapolation
should be performed first, and the flow-time-to-zero extrapolation should be
carried out on the continuum correlation functions. 
Otherwise, lattice-spacing issues which are ameliorated by flow reemerge in the flow-time-to-zero extrapolation.

The continuum extrapolations should only be performed at the separations of the finest lattice that is available, as the information is most reliable there. 
All data from coarser lattices need to be interpolated to these separations, for which we use cubic splines.
For a given lattice and flow time we only interpolate inside the interval given by \cref{flow_time_full_limit}, with the exception of the one data point that lies just outside of the lower separation limit. 
By also including this slightly contaminated point we can always interpolate at the lower limit without explicitly using the contaminated data, which extends the usable flow range. 
At the left boundary of the spline the second derivative is set to zero (natural condition) and at the middle point ($\tau T=0.5$, which in practice is the right boundary) the first derivative is set to zero (symmetry).
We perform the interpolations on every bootstrap sample and take the bootstrap mean and standard deviation of the interpolated points for statistical error estimation. 

Once the interpolated values are obtained, the continuum extrapolation is carried out, independently for each distance $\tau$ and flow time $\tf$, by performing a weighted fit on the data with the \textit{Ansatz} 
\begin{align}
    \frac{G^\textrm{latt}_{\tau,\tf}(N_\tau) }{G_{\tau,\tf=0}^{\begin{subarray}{l}\mathrlap{\textrm{\tiny  norm}}\\[-0.4ex] \textrm{\tiny latt}\end{subarray}}(N_\tau)}
     =  m \cdot N_\tau^{-2} + b 
\end{align}
with fit parameters $m$ and $b = G_{\tau,\tf}^\mathrm{cont}/G_{\tau,\tf=0}^{\begin{subarray}{l}\mathrlap{\textrm{\tiny  norm}}\\[-0.4ex] \textrm{\tiny cont}\end{subarray}}$. 

\begin{figure}[!t]
    \includegraphics[width=\figurewidth]{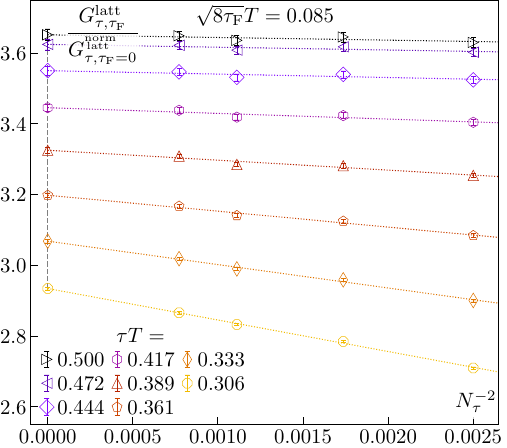}
    \caption{Continuum extrapolation of the color-electric correlator at $\sqrt{8\tf}T=0.085$ using an \textit{Ansatz} linear in $N_\tau^{-2}$. The lattices have temporal extents $N_\tau \in \lbrace 20,24,30,36 \rbrace$.}
    \label{fig:EE_cont_extr}
\end{figure}

The Wilson action that we use to generate the gauge configurations has leading discretization errors of order $a^2$, which explains this choice. 
We estimate the statistical error by performing the weighted fit on substitute bootstrap samples that we draw from a Gaussian around the bootstrap mean of the interpolations with their error as the width. 
The bootstrap mean and standard deviation of the substitute samples then serve as the statistical estimates.
In \autoref{fig:EE_cont_extr} we show the continuum extrapolation for one intermediate flow time ($\sqrt{8\tf}T = 0.085$).

\subsection{Flow-time-to-zero extrapolation}
\label{sec:zero-flow-extr}

The continuum correlators at fixed flow time still need to be extrapolated back to $\tf =0$. Because of the limited knowledge we have of the functional form of the flow effect we use the simple linear \textit{Ansatz}
\begin{align}
    \frac
    {G_{\tau}^\mathrm{cont}(\tf)}
    {G_{\tau,\tf=0}^\stackss{norm}{cont}}
    = m \cdot \tf + b,
\end{align}
where $m$ and $b=G^\mathrm{cont}_\tau/G_{\tau}^\stackss{norm}{cont}$ are $\tau$-dependent fit parameters. For statistical error estimation we proceed as in the continuum extrapolation. 

\begin{figure}[!t]
    \centering
    \includegraphics[width=\figurewidth]{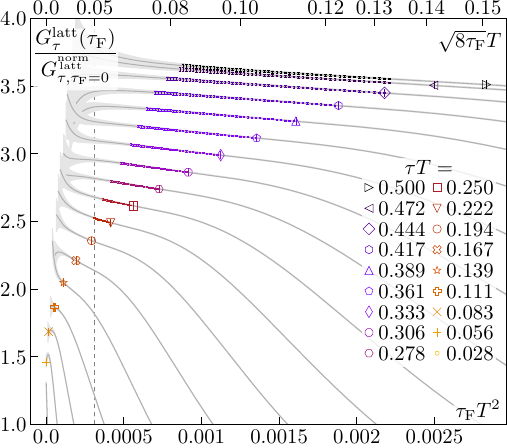}
    \caption{The color-electric correlator as a function of flow time on the finest available lattice ($N_\tau =36$) at $T\approx 1.5 \,T_c$. The grey error bands indicate the underlying data; to reduce clutter they are hidden if statistical errors would be larger than $5\%$. In retrospect we only consider the flow times for the flow extrapolation where the data points are shown explicitly (see \autoref{sec:zero-flow-extr}). The dashed vertical line is the minimum amount of flow that is needed for the coarsest considered lattice ($N_\tau=20$, see \autoref{tab:lattice_setup}). The horizontal positions of the various markers depict the flow limits from \autoref{flow_time_full_limit} (with $a = 1/(N_\tau T)$, $N_\tau = 20$).}
    \label{fig:EE_flow_effect}
\end{figure}

\begin{figure}[!t]
    \includegraphics[width=\figurewidth]{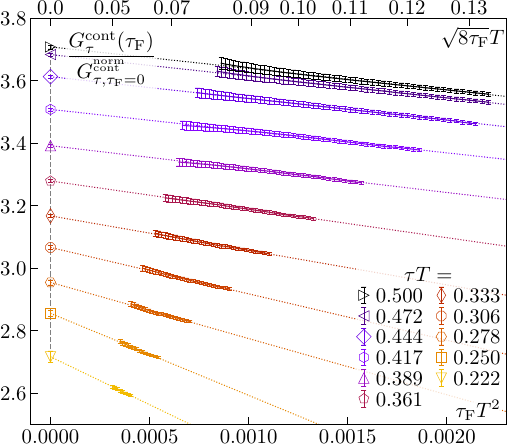}
    \caption{Flow-time-to-zero extrapolation of the continuum color-electric correlator at fixed $\tau T$ using a linear \textit{Ansatz}. The range of flow times at each separation is restricted by \autoref{flow_time_full_limit} and the statistical precision (see \autoref{sec:zero-flow-extr}).}
    \label{fig:EE_flow_extr}
\end{figure}

Based on the continuum correlators we estimate by hand for which separations and flow times this \textit{Ansatz} is reasonable (see \autoref{fig:EE_flow_extr} and also \autoref{fig:EE_flow_effect}):\\
For the lower boundary of the flow time the criterion is the statistical precision of the data. 
There are two ways to improve the precision: 1. apply more flow, 2. increase the number of gauge configurations. 
For the larger distances where one needs more configurations for the same statistical accuracy (in comparison to the smaller distances), the minimum amount of flow from \cref{flow_time_requirement} is not enough to give sufficiently small statistical errors on the data. 

Therefore we only use continuum data points in the flow-time-to-zero extrapolation that are flowed enough so that they have a maximum relative error of $1\%\cdot\tau T$. This value is chosen to yield a lower boundary for the flow time where a linear \textit{Ansatz} is reasonable. 
This also implies that the minimal separation for which we can obtain renormalized correlator data is shifted.
For the upper boundary of the flow time we are limited by \cref{flow_time_full_limit} and by the number of data points of the coarsest lattice; 
if there are less than two noncontaminated data points for any given lattice, we cannot perform the interpolation and continuum extrapolation. 
For the coarsest lattice this turns out to be at a flow time of $\sqrt{8\tf}T > 0.133$.

With all of these limitations we can reliably extrapolate the renormalized continuum correlator in the range $\tau T \in [8/36,0.5]$.  
In \autoref{fig:EE_final_comparison} we show the final renormalized continuum correlator at zero flow time. 
The errors shown here only depict statistical uncertainties. We also carry out a new continuum extrapolation of the lattice correlators from a previous study \cite{Francis:2015daa} which used the multilevel algorithm and link integration techniques to obtain a signal for the color-electric correlator. 
The old method only works in pure gauge theory and needs to be renormalized approximately with the help of perturbation theory \cite{Christensen:2016wdo}. 
Note that the statistics used in \cite{Francis:2015daa} are much smaller than the ones in this study (\autoref{tab:lattice_setup}), especially for the finer lattices. 

The two results agree in their overall shape, while the slight offset can be explained by missing higher-order contributions to the renormalization of the multilevel result, the systematic uncertainty introduced by the flow extrapolation, and the difference in lattices and statistics.

A note on systematic uncertainties: 
In principle, the correlator data shown in \autoref{fig:EE_final_comparison} is subject to systematic uncertainties through scale setting, auto-correlation between gauge configurations and cross-correlation between the correlator distances, the nonflowed tree-level improvement of the Zeuthen-flowed correlators, and finally the interpolation and extrapolation \textit{Ansätze}.

\section{Heavy quark momentum diffusion coefficient}\label{sec:hq-mom-diff}

\subsection{Spectral function of the color-electric correlator}

It was shown in \cite{CaronHuot:2009uh} that the heavy quark momentum diffusion coefficient can be obtained from
\begin{equation}
    \label{definition-kappa}
    \kappa = \lim_{\omega \rightarrow 0} \frac{2 T}{\omega} \rho(\omega)\,,
\end{equation}
where $\rho(\omega)$  is the spectral function of the color-electric correlator,
which is related to the Euclidean correlator we compute here through the integral relation
\begin{align}
    G(\tau) = \int_0^\infty \frac{{\rm d}\omega}{\pi} \rho(\omega)\frac{\cosh \big{(}\omega (\tau-\frac{1}{2T}) \big{)} } {\sinh \big{(} \frac{\omega}{2T} \big{)} }.
    \label{relation}
\end{align}
In the following we will calculate the leading-order continuum spectral function under gradient flow, which will lead to a discussion about the Kramers-Kronig relation and flowed retarded correlators. 

\begin{figure}[!t]
    \centering
    \includegraphics[width=\figurewidth]{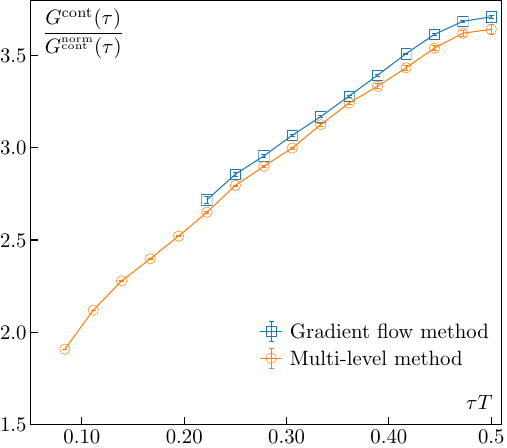}
    \caption{Nonperturbatively renormalized continuum color-electric correlator at zero flow time obtained from the gradient flow method (with lattice setup from \autoref{tab:lattice_setup}) in comparison with revised continuum correlator from multilevel method (perturbatively renormalized to NLO \cite{Christensen:2016wdo}) using lattice data from \cite{Francis:2015daa}.}
    \label{fig:EE_final_comparison}
\end{figure}

\begin{figure*}[t!]
    \centering{
    \includegraphics[width=\figurewidth]{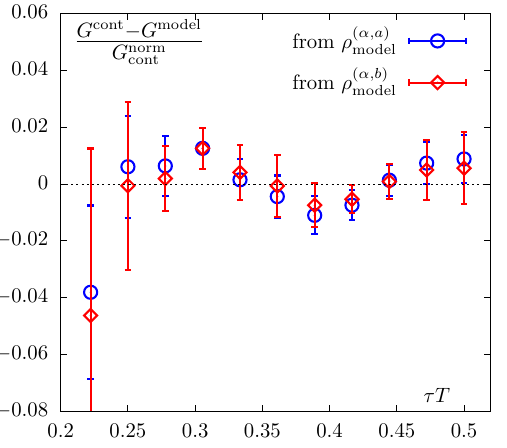}
    \hspace{1.5cm}
    \includegraphics[width=\figurewidth]{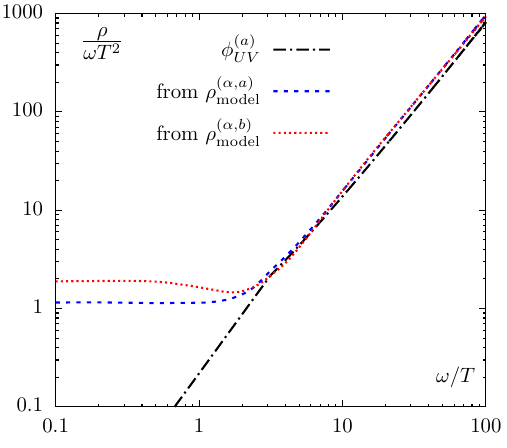}}
    \caption{Left: the differences between
the double-extrapolated correlators and the fit correlators for two models. Right: the fit spectral functions from different models (with $n_\text{max}=4$ in strategy I).}
    \label{fig:fitted-corrs}
\end{figure*}

The (continuum, zero flow time) leading-order spectral function is well known~\cite{Burnier:2010rp}.
But we did not find a study of the spectral function for the Euclidean
function at finite flow depth, so we will consider this problem here.
We start from the leading-order correlation function at flow time $\tf$, 
\begin{align}
    \begin{split}
        G^{\textrm{\tiny LO}}(\tau,\tf)=&-\frac{\gb^{2} \CF}{3} \sumint\limits_{K} e^{i k_{n} \tau} e^{- 2\tf K^2} \frac{(D-1)k_{n}^{2}+k^{2}}{K^{2}} \\
        =&-\frac{\gb^{2} \CF }{3} \frac{d}{(4 \pi)^{d/2}} T\sum_{k_n} e^{i k_{n} \tau} \left(k_{n}^{2}\right)^{d/2} \\
        &\times \Big{(} \Gamma(1-d/2,2 \tf k_{n}^{2})+ \frac{1}{2} \Gamma(-d/2,2 \tf k_{n}^{2}) \Big{)} \,,
    \end{split}
\end{align}
where $\Gamma(s,x)= \int_{x}^{\infty}\diff t~ t^{s-1} e^{-t}$ is the incomplete Gamma function and $d$ the number of space dimensions.
Fourier transforming the correlator to frequency space using the (periodic) Kronecker delta function $\int_{0}^{\beta}\diff \tau e^{i k_{n} \tau}= \beta \delta_{k_{n},0}$ leads to 
\begin{equation}
    \begin{aligned}
        \tilde{G}^{\textrm{\tiny  LO}}(\omega_{n},\tf)=&-\frac{\gb^{2} \CF }{3} \frac{d}{(4 \pi)^{d/2}} \left(\omega_{n}^{2}\right)^{d/2}
        \\
        & \times  \Big{(} \Gamma(1-d/2,2 \tf \omega_{n}^{2})+ \frac{1}{2} \Gamma(-d/2,2 \tf \omega_{n}^{2}) \Big{)}\,.
    \end{aligned}
\end{equation}
Setting the dimension to its physical value $d=3$ and performing the analytic continuation 
\begin{align}
    \label{eq: analytic continuation}
    \rho(\omega)= \Imp\Big{[} \lim_{\epsilon \rightarrow 0^{+}} \tilde{G}(\omega_{n}\rightarrow -i \omega +\epsilon) \Big{]}\,,
\end{align}
we find the spectral function to be
\begin{align}
    \label{eq: EE flowed spectral function}
        \rho^{\textrm{\tiny LO}}(\omega,\tf)=&- \gb^{2} \CF \frac{\omega^{3}}{(4 \pi)^{3/2}}
        \\ \nonumber
        & \times \lim_{\epsilon \rightarrow 0^{+}} \Rep\Big{[} \Gamma\big{(}1-3/2,2 \tf (-i\omega+\epsilon)^{2}\big{)}
        \\ \nonumber
        & \qquad \qquad + \frac{1}{2} \Gamma\big{(}-3/2,2 \tf (-i\omega+\epsilon)^{2}\big{)} \Big{]} \,.
\end{align}
The real parts of the incomplete Gamma functions do not vary with their incomplete argument. 
The value of the real part of the incomplete Gamma function is therefore the same as the ordinary Gamma function. 
The leading order spectral function is 
 \begin{align}
    \label{eq: EE flowed spectral function result d=3/2}
    \rho^{\textrm{\tiny  LO}}(\omega,\tf)=\frac{\gb^{2} \CF}{6 \pi}\omega^{3}\,,
\end{align}
which is the same result as the nonflowed spectral function calculated in \cite{Burnier:2010rp}.
Therefore, at leading perturbative order, the spectral function for the
flowed and for the nonflowed Euclidean correlation function are the same.
At first this appears remarkable; two distinct Euclidean functions reconstruct
to the same spectral function, which implies that \cref{relation} must
not apply to the flowed correlation function.

The connection between the retarded correlator and the spectral function is made by the Kramers-Kronig relation (see \cite{Jackson:1998nia}). 
The real and imaginary parts of a meromorphic function, which is analytic in the closed upper half-plane, are related via Kramers-Kronig integral formulas if the function vanishes fast enough for a growing complex argument. 
The mathematical requirements of the function can be translated to the necessity of causality of the underlying physical theory. 

The introduction of gradient flow, that is, replacing local operators with smeared-out relatives, breaks energy-momentum conservation through new contact terms. 
The commutator correlator of smeared-out operators no longer vanishes outside the light cone, thus breaking causality.  Alternatively, we can appeal to
\cref{flow_solution}, where $e^{-\tf p^2}$ represents an exponential
suppression for Euclidean 4-momentum $p$, but is an exponential enhancement
for timelike Minkowski 4-momenta.
Mathematically speaking, the exponential suppression along the Euclidean time axis $\tau$ will turn to exponential growth after Wick rotation. 
This happens during the analytic continuation of the Euclidean to the retarded correlator, which no longer fulfills the requirements needed for the Kramers-Kronig relation, thereby breaking its connection to the spectral function.

This discussion proves that spectral reconstruction can only be attempted
on data which has been extrapolated to zero flow time (as well as to the
continuum). 
As we have already emphasized, this extrapolation should be performed \textit{after} extrapolating to the continuum limit.
With the extrapolated data we can apply spectral reconstruction techniques to estimate the heavy quark momentum diffusion coefficient $\kappa$, which we will do in the next section.

\subsection{Extraction of the heavy quark momentum diffusion coefficient}

There are many approaches for a spectral reconstruction~\cite{Tikhonov:1620560,PhysRevE.95.061302,JARRELL1996133,Burnier:2013nla,Ding:2017std,10.1111/j.1365-246X.1968.tb00216.x}. 
Here we will proceed similarly to \cite{Francis:2015daa} and reconstruct the spectral function based on theoretically motivated model fits of the correlator data obtained from the gradient flow method that is shown in \autoref{fig:EE_final_comparison}. 
The fit ansatz is a combination of two parts: one is valid in the small frequency regime ($\omega \ll T$) and the other is valid in the large frequency regime ($\omega \gg T$). 
In the intermediate regime an interpolation is required. 
The small frequency part can be described by infrared asymptotics~\cite{CaronHuot:2009uh},
\begin{align}
    \phi_{\rm{IR}}(\omega) \equiv \frac{\kappa \omega}{2T},
    \label{phiIR}   
\end{align}
and the large frequency part can be calculated in perturbation theory with the help of asymptotic freedom~\cite{CaronHuot:2009ns,Burnier:2010rp},
\begin{align}
    \phi^{(a)}_{\rm{UV}}(\omega) \equiv \frac{g^2(\bar{\mu}_\omega) C_F \omega^3}{6\pi}
     \;, \quad
     \bar{\mu}_\omega \equiv \mbox{max}(\omega,\pi T),
    \label{phiUV1}
\end{align}
or, when considering the higher order correction,
\begin{align}
    \phi^{(b)}_{\rm{UV}}(\omega) \equiv \phi^{(a)}_{\rm{UV}}(\omega) \Big{[} 1 + (r_{20} + r_{21} \ell) a_s (\bar{\mu}_\omega) \Big{]}.
    \label{phiUV2}
\end{align}
$C_F$, $r_{20}$, $r_{21}$, $\ell$ can be found in~\cite{Burnier:2010rp, Francis:2015daa} and $g^2$, $a_s$ are evaluated using four-loop
running~\cite{vanRitbergen:1997va}. Equation (\ref{phiUV1}) and \cref{phiUV2} will be treated equally in our fits as explained in \cite{Francis:2015daa}. 
To account for the corrections when incorporating $\phi_{\rm{IR}}$ and $\phi_{\rm{UV}}$, which are valid only in their own regimes, we need to introduce two trigonometric functions,
\begin{align}
    e^{(\alpha)}_n(y) \equiv \sin(\pi n y),\ \ e^{(\beta)}_n(y) \equiv \sin(\pi y) \sin(\pi n y),
    \label{e_alpha_beta}
\end{align}
where $y \equiv \frac{x}{1+x}$ and $x \equiv \ln\Big{(} 1 + \frac{\omega}{\pi T} \Big{)}$. Putting all this together yields three types of models as in \cite{Francis:2015daa}; however, in this work we only adopt the most theoretically justified one,
\begin{equation} 
     \rho_{\rm{model}}^{(\mu, i)}(\omega) \equiv \Big{[} 1 + \sum_{n=1}^{n_{\rm{max}}} c_n e^{(\mu)}_n(y) \Big{]} \sqrt{ \big{[} \phi_{\rm{IR}}(\omega)\big{]}^2 + \big{[} \phi^{(i)}_{\rm{UV}}(\omega)\big{]}^2 },
\end{equation}
where $\mu \in \{\alpha,\beta \}$, $i \in \{ a,b\}$. Substituting the true spectral function in \cref{relation} with the model spectral function, we obtain a model correlator $G^\textrm{model}$ with which we define a weighted sum of squares for a fit:
\begin{align}
    \chi^2\equiv\sum_{\tau}\bigg{[}\frac{G^{\rm{cont}}(\tau)-G^\textrm{model}(\tau)}{\delta G^{\rm{cont}}(\tau)}\bigg{]}^2.
\end{align}
Here $\delta G^{\rm{cont}}(\tau)$ is the statistical error of the lattice data $G^{\rm{cont}}(\tau)$ which we show in \autoref{fig:EE_final_comparison}.

Now it is clear that the parameters to fit are $\kappa/T^3$ and $c_n$. 
As in \cite{Francis:2015daa}, two strategies are considered. 
The first strategy is just a normal fit and in the second strategy we constrain that at the maximum frequency the spectral function reproduces its UV asymptotics. 
In strategy I we do not stop after 200 iterations as in~\cite{Francis:2015daa} as we have seen that this criterion makes almost no difference. 
For both strategies in our fit routine we set the absolute tolerance for $\chi^2$ to 0.0001 and the maximum number of iterations to 2000. 
In our fits we limit ourselves to $n_\text{max}\in\lbrace 4,5\rbrace$ and we use all the data points shown in \autoref{fig:EE_final_comparison}, that is, $\tau T\ge 8/36$.

\begin{figure}[t!]
    \centering
    \includegraphics[width=\figurewidth]{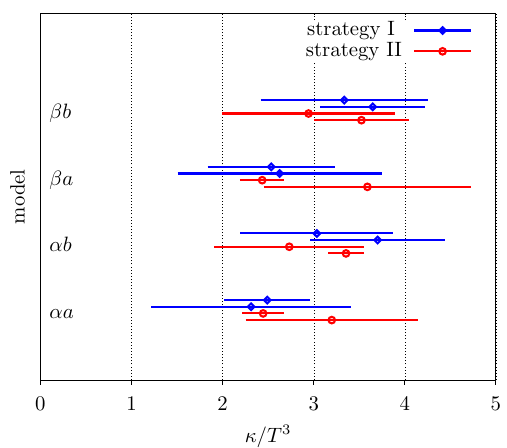}
    \caption{$\kappa/T^3$ from different fit models and strategies. For each model the lower data point corresponds to $n_\text{max}=4$, while the upper one corresponds to $n_\text{max}=5$.}
    \label{fig:kappa}
\end{figure}

In \autoref{fig:fitted-corrs} we show an example of the fit correlators and spectral functions when using model $\rho_{\rm{model}}^{(\alpha,a)}$ and $\rho_{\rm{model}}^{(\alpha,b)}$ with $n_\text{max}=4$ in strategy I. 
The errors shown in the left figure are statistical errors obtained from a bootstrap analysis. 
In the right figure the error bars are not shown to gain better visibility. 
We see that the two model spectral functions describe our data in a similar way as the differences between the data and fit correlators are also similar. 
This also holds for all the other choices of models or fit strategies and we obtained $\chi^2/\text{d.o.f.}\sim 3.4\,\dots\,5.5$ throughout. 
The resultant $\kappa/T^3$ is summarized in \autoref{fig:kappa}. 
For each point the error bar is the statistical uncertainty we obtain from a bootstrap analysis and we take the spreading over various models as the systematic uncertainty of our estimation. 
Finally, based on the central values of \autoref{fig:kappa} we obtain a range for $\kappa/T^3$:
\begin{equation}
    \kappa/T^3=2.31\,\dots\,3.70,
\end{equation}
which is slightly shifted to larger values compared to $1.8\,\dots\,3.4$ from \cite{Francis:2015daa} and $1.31\,\dots\,3.64$ from \cite{Brambilla:2020siz}. 
With the estimated range for $\kappa/T^3$ we can also estimate the heavy quark diffusion coefficient in the heavy quark mass limit ($M\gg \pi T$) according to $D=2T^2/\kappa$~\cite{CaronHuot:2009uh}:
\begin{equation}
    DT=0.54\,\dots\,0.87.
\end{equation}

\section{Conclusion}\label{sec:conclusion}
We calculated the color-electric correlator on four isotropic lattices at $T\approx 1.5\,T_c$ in the quenched approximation under gradient flow. 
We found that the gradient flow technique significantly improves the signal of the correlator on the lattice. 
It is necessary to perform both a continuum and a flow-time-to-zero extrapolation; we show that the continuum
limit must be taken first, and we perform both
extrapolations on our data
within the range $\tau T \in [8/36,0.5]$. 
We found that an \textit{Ansatz} linear in $1/N_{\tau}^2$ and one linear in $\tau_F$ is suitable for the respective extrapolations. 
The overall shape of the correlator obtained via the flow method in the $a\rightarrow 0$ and $\tau_F\rightarrow 0$ limit largely agrees with its multilevel counterpart in the $a\rightarrow 0$ limit. The slight offset can be explained by missing higher-order contributions to the renormalization of the multilevel result, the systematic uncertainty introduced by the flow extrapolation, and the difference in lattices and statistics.
The heavy quark diffusion coefficient obtained from the double-extrapolated correlator is also consistent with but slightly larger than that from the correlator computed with the multilevel approach, according to the $\chi^2$ fits performed on the correlators using theoretically well-established models. 

This work affirms the validity of the gradient flow technique when performing the continuum and flow-time-to-zero extrapolations and provides a method for extending these studies to full QCD. 

All data from our calculations, presented in the figures of this paper, can be found in \cite{datapublication}.

\begin{acknowledgments}
We would like to thank Mikko Laine for useful discussions and previous collaboration. The authors acknowledge support by the Deutsche Forschungsgemeinschaft (DFG, German Research Foundation) through the CRC-TR 211 'Strong-interaction matter under extreme conditions'– project number 315477589 – TRR 211. The computations in this work were performed on the GPU cluster at Bielefeld University.
\end{acknowledgments}

\section*{Appendix}
\appendix

\section{Tree-level improvement and flow}
\label{sec:appendix-tree-level-flow}

\begin{figure*}[!t]
    \hspace*{\fill}
    \includegraphics[width=\figurewidth]{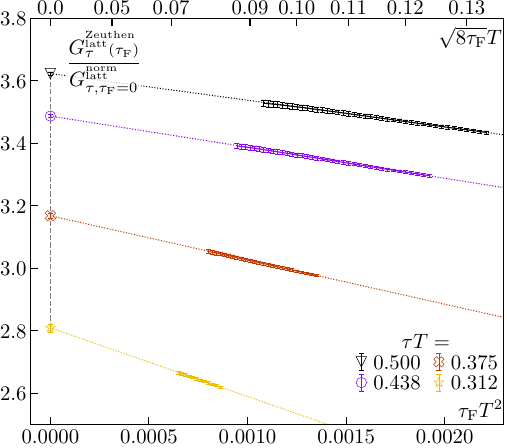}
    \hfill
    \includegraphics[width=\figurewidth]{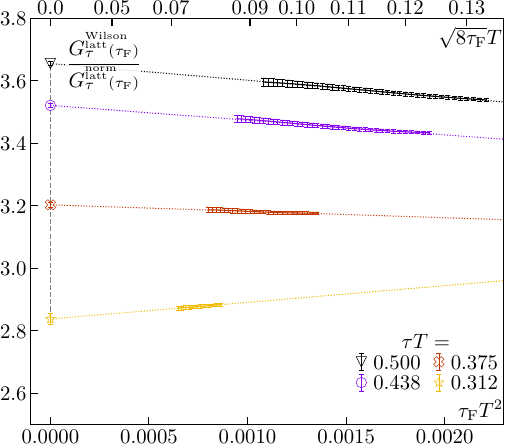}
    \hspace*{\fill}
    \caption{Left: naive flow-time-to-zero extrapolation of the \textit{Zeuthen-flowed} color-electric correlator on a $64^3 \times 16$ lattice at $T\approx 1.5\,T_c$. The correlator has been tree-level-improved by using the \textit{nonflowed} leading-order continuum and lattice correlators (as in \autoref{Gimp_zeuthen}).
    Right: \textit{Wilson-flowed} correlator in the same setting. The correlator has been tree-level-improved by using the \textit{Wilson-flowed} leading-order continuum and lattice correlators at the same flow time.}
    \label{fig:EE-Nt16-flow-extr}
\end{figure*}

\subsection{Derivation of leading-order lattice correlator under Wilson flow}\label{subsec:appendix-lo-derivation}

In this section we present the derivation of the leading-order lattice correlator under Wilson flow. 
The lattice gluon propagator under Wilson flow has the form
\begin{align*}
    &\left\langle B^{a}_{\mu}(K,\tf) B^{b}_{\nu}(Q,\tf) \right\rangle_{0}=\delta^{ab} ~(2\pi)^{4}\delta^{(4)}(K+Q)
    \\
    &\times  \frac{1}{\tilde{K}^{2}} \left[ \left( \delta_{\mu \nu} -  \frac{\tilde{K}_{\mu} \tilde{K}_{\nu}}{\tilde{K}^{2}} \right) e^{-2\tf \tilde{K}^{2}}  +\xi \frac{\tilde{K}_{\mu} \tilde{K}_{ \nu}}{\tilde{K}^{2}} e^{-\alpha_{0} 2 \tf \tilde{K}^{2}} \right]\,,
\end{align*}
where $\xi$ and $\alpha_{0}$ are the gauge and flow gauge fixing parameters and we used the tilde short-hand notation
\begin{equation}
\label{eq: LPT shorthand}
    \tilde{k}_{\mu}=\frac{2}{a} \sin \left(\frac{a k_{\mu}}{2} \right)\,.
\end{equation}{}
In the following we will use both Feynman and flow Feynman gauge. 

The leading-order expression reads
\begin{align}
    \label{eq: leading order lattice}
    G^{\textrm{\tiny LO}}_{\textrm{\tiny latt}}&(\tau,\tf) = -\frac{\gb^{2}\CF}{3 } \sumint\limits_{K} e^{i  k_{n} \tau} e^{- 2\tf \tilde{K}^{2}} \left[ 3- 2 \frac{\sum_{i} \tilde{k}_{i}^{2}}{\tilde{K}^{2}} \right]\,,
\end{align}
which is analogous to the continuum expression but exchanging $\tilde{k} \rightarrow k$. The spatial integration can be performed analytically 
\begin{align*}
    &\int\limits_{-\pi}^{\pi}\frac{d^{3}k}{(2 \pi)^{3}} e^{- 8 \frac{\tf }{a^{2}} \sum\limits_{i}  \sin^{2}\left( \frac{ k_{i}}{2} \right)} = e^{-12 \frac{\tf }{a^{2}} } \left( I_{0}\left(4 \frac{\tf }{a^{2}}\right)\right)^{3}\,,
    \\
    &  \int\limits_{-\pi}^{\pi}\frac{d^{3}k}{(2 \pi)^{3}} \sum\limits_{i}  \sin^{2}\left( \frac{ k_{i}}{2} \right) e^{- x  \sum\limits_{i}  \sin^{2}\left( \frac{ k_{i}}{2} \right) }  
    \\
    & \hspace{3.0em}= \frac{3}{2} e^{-\frac{3x}{2}} \left( I_{0}\left( \frac{x}{2}\right)\right)^{2} \left\lbrace   I_{0}\left( \frac{x}{2}\right) - I_{1}\left( \frac{x}{2}\right) \right\rbrace\,,
\end{align*} 
where $I_{0}$ and $I_{1}$ are modified Bessel functions of the first kind. Inserting this into \cref{eq: leading order lattice} and introducing a Schwinger parameter in order to rewrite the denominator leads to the expression given in \cref{glatflow1}.

\subsection{Flow-time-to-zero extrapolation and comparison of tree-level improvement methods under flow}\label{sec:tree-level-imp}

Correlators calculated on the lattice suffer from cutoff effects. One can extract and account for the leading-order contribution of these effects by comparing the perturbative continuum and lattice correlator (``tree-level improvement'' \cite{Sommer:1993ce}). This can be achieved by defining the improved distances $\overline{\tau T}$ through 
\begin{align}
    G_\tf^\stackss{norm}{cont}(\overline{\tau T})
    &\equiv 
    G_\tf^\stackss{norm}{latt}(\tau T)
\end{align}
which are used to shift the correlator values according to
\begin{align}
    G_\tf^{\text{imp}}(\overline{\tau T})
     &\equiv 
     G_\tf^{\mathrm{latt}}(\tau T).
\end{align}

Another possibility is given by multiplying the correlator data at their original distances with ratio of the leading-order continuum and lattice correlator:
\begin{align}
    \label{eq:impfactor}
    G_\tf^{\text{imp}}(\tau T)
    \equiv
    \frac
    {G_\tf^\stackss{norm}{cont}(\tauT)}
    {G_\tf^\stackss{norm}{latt}(\tau T)}
    G^\textrm{latt}_\tf(\tau T).
\end{align}

In principle it should not matter whether one improves the distances or the correlator values themselves. 
However, the interpolations of the correlator data that are necessary for the continuum extrapolation will change slightly. 
The data will not be distributed evenly if one improves the separations, and one obtains slightly smaller large separations (e.g. $\tau T=0.5 \rightarrow \overline{\tau T} \approx 0.48$) which is unwished-for since most of the information about the low-frequency part of the spectral function is contained at the largest $\tau T$. 
Consequently, we choose the straightforward strategy of \cref{eq:impfactor} to perform the tree-level improvement.

In the end we want to normalize our lattice  measurements to the leading-order continuum correlator and so we end up with 
\begin{align}
    \label{tree-level-imp}
    \frac
    {G_\tf^\textrm{imp}(\tauT)}
    {G_\tf^\stackss{norm}{cont}(\tauT)}
    = 
    \frac
    {G_\tf^\stackss{norm}{cont}(\tauT)}
    {G_\tf^\stackss{norm}{latt}(\tauT)}
    \frac
    {G_\tf^\mathrm{latt }(\tauT)}
    {G_\tf^\stackss{norm}{cont}(\tauT)}
    = 
    \frac
    {G_\tf^\mathrm{latt}(\tauT)}
    {G_\tf^\stackss{norm}{latt}(\tauT)}.
\end{align}
In principle it is necessary to take both correlators on the right side of \cref{tree-level-imp} at the same flow time $\tf$; however, currently we have only calculated the leading-order color-electric correlator under Wilson flow [\cref{eq: leading order lattice}], which is substantially different from the more complicated Zeuthen flow~\cite{Ramos:2015baa} that we use for the numerical simulations. 

\begin{figure}[!t]
    \centering
    \includegraphics[width=\figurewidth]{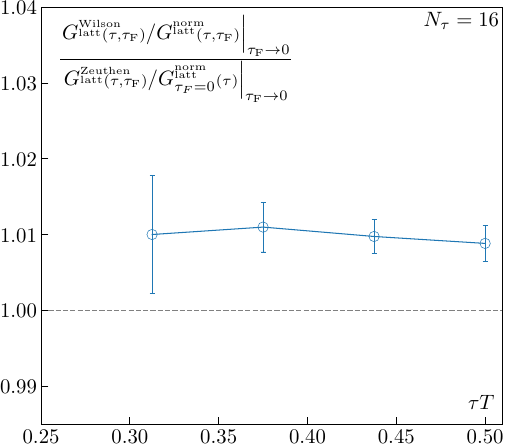}
    \caption{Ratio of the Zeuthen-flowed correlator with nonflowed tree-level improvement and the Wilson-flowed correlator with Wilson-flowed tree-level improvement on the $64^3\times 16$ lattice for large $\tau T$.}
    \label{fig:EE-flow-imp-comparison}
\end{figure}

In \autoref{fig:EE-Nt16-flow-extr} we compare the \textit{Zeuthen-flowed} correlator improved by the \textit{nonflowed} leading-order correlators (left panel) with the \textit{Wilson-flowed} correlator improved by the  \textit{Wilson-flowed} leading-order correlators at the same flow time (right panel) on a coarse lattice with $N_\tau =16$. 
While the behavior under flow of these two quantities differs greatly, their naive flow-time-to-zero extrapolations using a linear \textit{Ansatz} (see also \autoref{sec:zero-flow-extr}) differ only by about $1\%$, which is shown in \autoref{fig:EE-flow-imp-comparison}. 
This difference is mainly caused by the discretization, and we expect it to be even smaller for finer lattices, which is why we decide to use the nonflowed tree-level improvement on the Zeuthen-flowed measurements:
\begin{align}
    \label{Gimp_zeuthen}
    G^\textrm{imp}_\tf(\tauT)
    \equiv 
    \frac
    {G_{\tf=0}^\stackss{norm}{cont}(\tauT)}
    {G_{\tf=0}^\stackss{norm}{latt}(\tauT)}
    G^{\begin{subarray}{l}\mathrlap{\textrm{\tiny  Zeuthen}}\\[-0.3ex] \textrm{\tiny latt}\end{subarray}}_{\tf} {(\tauT)}
    .
\end{align}
The continuum and flow-time-to-zero extrapolations are performed on this improved correlator normalized by the leading-order continuum correlator.

\section{Electric field renormalization under Wilson flow}\label{sec:appendix-renorm-flow}

\begin{figure}[t!]
    \centering
    \includegraphics[width=\figurewidth]{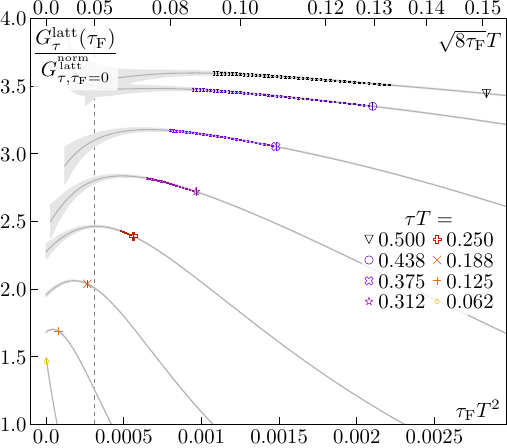}
    \caption{Nonperturbative Wilson-flowed color-electric correlator with Wilson-flowed tree-level improvement as a function of flow time on a $64^3\times 16$ lattice at $1.5\,T_C$.}
    \label{fig:EE_Nt16_Wilson}
\end{figure}

In \autoref{sec:Cutoffeffects} we briefly comment on the initial rising behavior of the color-electric correlator which is visible for the second smallest distance and beyond if the statistical errors are small enough. 
(The rising behavior is not visible for the higher separations as here the dominant effect of the flow is the improvement of the signal.) \autoref{fig:EE_Nt16_Wilson} shows the Wilson-flowed (not Zeuthen) color-electric correlator as a function of flow time on a $N_\tau=16$ lattice where the effect is clearly visible. We expect this behavior to stem from the renormalization of the color-electric field which is caused by the flow.  
The purpose of this section is to give a qualitative explanation for this behavior. This is done by calculating the two-point function of the electric field in compact QED lattice perturbation theory in vacuum as a function of the flow time to next-to-leading order:
\begin{align}
    G_{\textrm{\tiny QED}}(\tau,\tf)=\left\langle \eb E_{i}(\tau,\vec{0},\tf) \eb E_{i}(0,\vec{0},\tf)\right\rangle\,.
\end{align}
This enables us to qualitatively analyze the renormalization behavior of the electric field under gradient flow with a smaller number of diagrams compared to QCD. Nonetheless, the renormalization properties should be the same in both theories. 

The leading order contribution has already been derived in the thermal theory (\autoref{eq: leading order lattice}) and so we find 
\begin{align}
    G^{\textrm{\tiny LO}}_{\textrm{\tiny QED}}(\tau,\tf)= \eb^{2} \int\limits_{K} e^{i k_{0} \tau} \frac{e^{- 2\tf \tilde{K}^{2}}}{\tilde{K}^{2}} \left[ 2 \tilde{k}_{0}^{2}+ \tilde{K}^{2} \right]\,.
\end{align}

\begin{figure}[!t]
    \hfill \includegraphics[width = 0.2\textwidth]{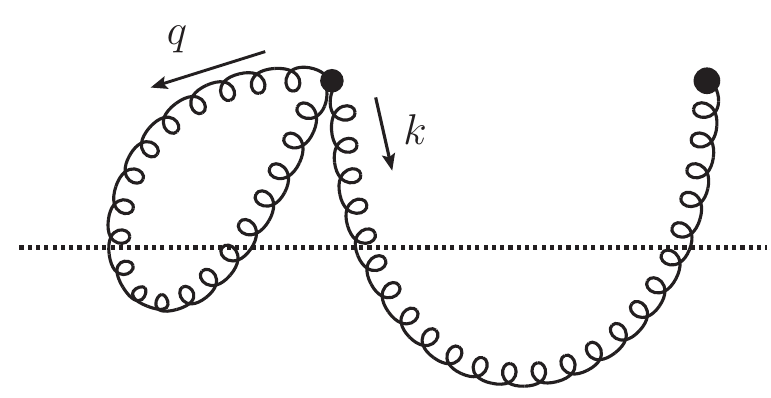}
    \hfill \includegraphics[width = 0.2\textwidth]{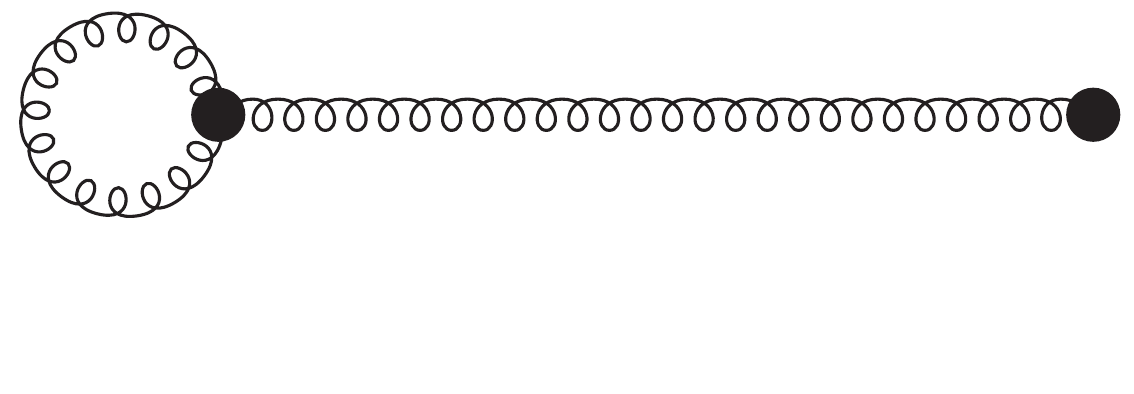}
    \hfill $\phantom{.}$
    \caption{Next-to-leading order diagram of the expansion of the color-electric field operator. On the left the diagram is drawn in our notation;
    below the horizontal dotted line are the nonflowed propagators and vertices,
    and the vertical axis above the dotted line represents evolution in flow time.
    On the right side the same diagram is drawn in the notation introduced in \cite{Luscher:2011bx}. We will denote this diagram to be contribution $(a)$.}
     \label{fig: NLO_diagram operator}
\end{figure}

\begin{figure}[!t]
    \hfill \includegraphics[width = 0.15\textwidth]{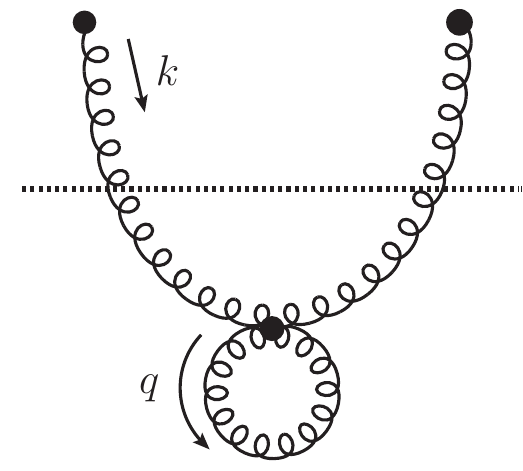}
    \hfill \includegraphics[width = 0.2\textwidth]{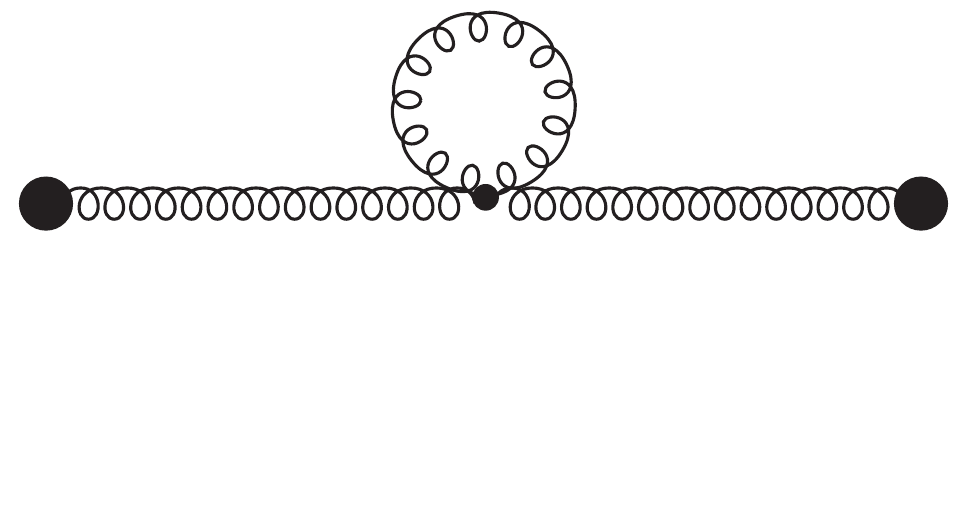}
    \hfill $\phantom{.}$
    \caption{\label{fig: NLO_diagram action}
    Next-to-leading order diagram of the expansion in the action. We call this diagram $(b)$.}
\end{figure}

\begin{figure}[!t]
    \hfill \includegraphics[width = 0.2\textwidth]{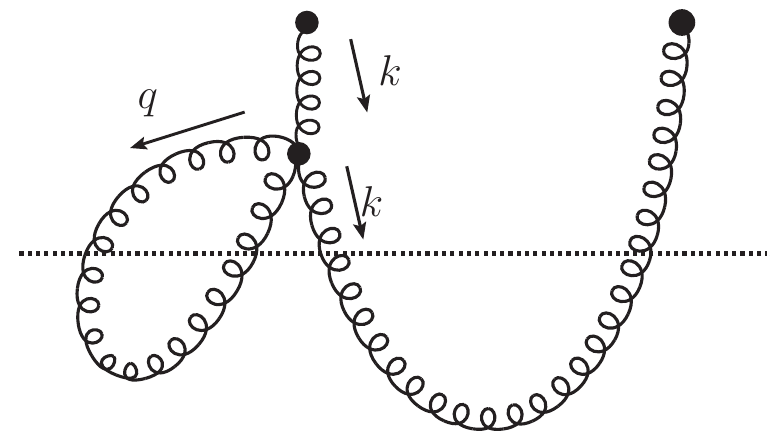}
    \hfill \includegraphics[width = 0.2\textwidth]{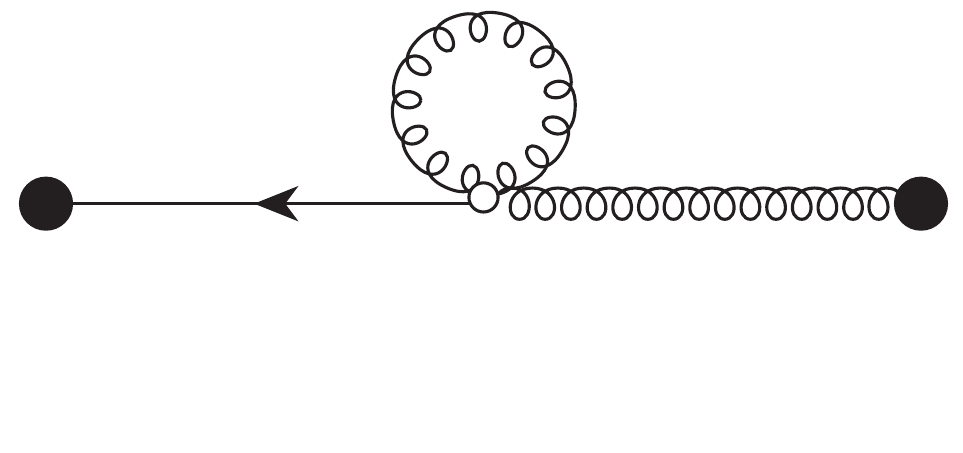}
    \hfill $\phantom{.}$
    \caption{\label{fig: NLO_diagram flow}
    Next-to-leading order diagram of the expansion of the flow equation. Only tree structures are possible for $\tf>0$. We call this diagram $(c)$.}
\end{figure}

At next-to-leading order in the lattice spacing $a$ there are three diagrams, which arise from the expansion of the electric field operator (\autoref{fig: NLO_diagram operator}), the action (\autoref{fig: NLO_diagram action}) and the Wilson flow equation (\autoref{fig: NLO_diagram flow}).

\begin{figure}[b!]
    \centering
    \includegraphics[width=\figurewidth]{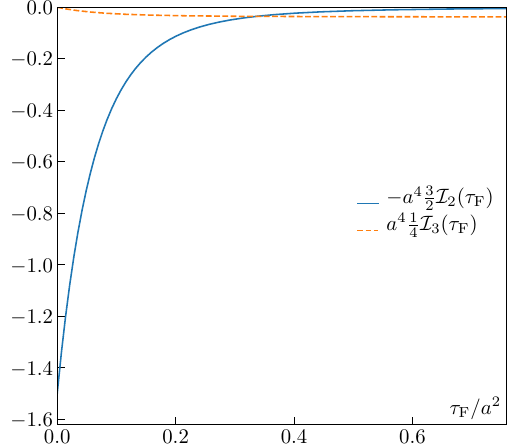}
    \caption{Two of the integral terms of the next-to-leading-order lattice perturbation theory calculation of the electric correlator in QED. The term involving $\mathcal{I}_{2}$ represents the tadpole renormalization as a function of flow time and is responsible for the initial rising behavior of the correlator. The term involving $\mathcal{I}_{3}$ is a pure Wilson flow effect and vanishes for zero flow time.}
    \label{fig:EE_LPT_QED}
\end{figure}

The tadpole diagram (\autoref{fig: NLO_diagram operator}) is a lattice artifact and describes the mixing between the electric field operator $E$ and higher-order operators such as $E^3$ which are induced by the lattice (lattice tadpole corrections).
The contribution of this unphysical type of diagram vanishes in the continuum limit as there is no equivalent diagram in the continuum. However, for noncontinuum-extrapolated calculations the contribution of the diagram needs to be considered and the interpretation is already well understood and we refer to~\cite{Capitani:2002mp} for a detailed analysis. One of the electric fields is expanded up to $\mathcal{O}(a^{4})$ leading to 
\begin{align}
    \delta_{(a)}G^{\textrm{\tiny NLO}}_{\textrm{\tiny QED}}(\tau,\tf) \hspace{-5.5em}& \nonumber
    \\
    =& \frac{-a^{4}\eb^{4}}{6}  \left.\left\langle \left(G_{0 i}(X,\tf) \right)^{3} G_{0 i}(X^{\prime},\tf) \right\rangle \right|_{X^{\prime}=0,  \vec{x}=0} \nonumber
    \\
    =& -a^{4}\eb^{4} ~ \text{LO}(\tau,\tf)  \int\limits_{Q} \frac{e^{- 2\tf \tilde{Q}^{2}}}{\tilde{Q}^{2}} \left[ 2 \tilde{q}_{0}^{2}+ \tilde{Q}^{2} \right] \nonumber
    \\
    =& -\frac{3}{2} a^{4}\eb^{4} ~ \text{LO}(\tau,\tf)  \underbrace{\int\limits_{Q} e^{- 2\tf \tilde{Q}^{2}}}_{\equiv \mathcal{I}_{2}(\tf)}
    \,.
\end{align} 
This diagram is a pure multiplicative correction to the leading order and describes the tadpole renormalization of the operators. The integral can be calculated analytically yielding
\begin{align}
    \mathcal{I}_{2}(\tf)=\int\limits_{Q} e^{- 2\tf \tilde{Q}^{2}}= e^{-16 \tf/a^{2}} \left(\frac{I_{0}(4 \tf/a^{2})}{a}\right)^{4}\,,
\end{align}
where  $I_{0}$ denotes the modified Bessel function of the first kind. The term $-3/2 \, a^{4}\, \mathcal{I}_{2}(\tf)$ is shown in \autoref{fig:EE_LPT_QED}. 
From the figure it is clear that for zero flow time the diagram has a significant correction to the leading-order behavior of the correlator. 
However, if more flow is applied, this contribution vanishes.
One can interpret this behavior as follows: the lattice introduces high-dimension operators with coefficients containing powers of $a$:  $E \to E + a^4 E^3$.  The flow smears out the fields and introduces a smooth momentum-cutoff, so that
the expectation value of such high-dimension operators is suppressed by an
inverse power of the flow time; the effects $a^4 E^3$ are parametrically
of order $a^4/\tf^2$.

Another NLO effect arises from interaction terms in the expansion of the action.
The lowest order interaction is a four-point vertex. The corresponding diagram is the first diagram contributing to the photon self-energy up to this order in $a$ and is shown in \autoref{fig: NLO_diagram action}.
 It has the same meaning as in the continuum: the renormalization of the coupling. Since the loop momentum lives at zero flow time, the corresponding correction is expected to be independent of the flow time.
Consequently, the correlator as a function of gradient flow takes the form
\begin{align}
    \delta_{(b)}G^{\textrm{\tiny NLO}}_{\textrm{\tiny QED}}(\tau,\tf) \hspace{-5.5em}&  \nonumber
    \\
    =& \frac{-a^{4}\eb^{4}}{48} \int\limits_{y} \left.\left\langle G_{0 i}(X,\tf) F^{4}_{\alpha \beta}(Y)  G_{0 i}(X^{\prime},\tf) \right\rangle \right|_{X^{\prime}=0, \vec{x}=0} \nonumber
    \\
    =& \frac{-a^{4}\eb^{4}}{4}  \text{LO}(\tau,\tf) \mathcal{I}_{1}
    \,,
\end{align}
where the integral $\mathcal{I}_{1}$ is defined as
\begin{align*}
    \mathcal{I}_{1}= \int\limits_{Q} 1 =a^{-4}\,.
\end{align*}
This shift is exactly the leading-order contribution to the difference between the bare and renormalized coupling.
 
The last diagram (\autoref{fig: NLO_diagram flow}) emerges from the nonlinear part of the flow equation leading to a four-point vertex at an intermediate flow time $\tau^{\prime}$. It is a pure gradient flow correction to the photon self-energy.
In QED this is a pure lattice artifact; in QCD similar terms also appear in the
continuum theory but the lattice introduces a distinct gauge-invariant subclass
of effects with no continuum analogue.
The unique NLO correction in lattice QED is
\begin{align}
    \delta_{(c)}G^{\textrm{\tiny NLO}}_{\textrm{\tiny QED}}(\tau,\tf) \hspace{-5.5em}& \nonumber
    \\
    =& \frac{-a^{4}\eb^{4}}{6} \int\limits_{Y} \int\limits_{0}^{\tf}d\tau^{\prime} \left.\left\langle G_{0 i}(X,\tf) \partial^{y}_{\alpha} L_{\beta}(Y,\tau^{\prime}) \right. \right. \nonumber
    \\
    & \hspace{8.0em} \left. \left. \times G^{3}_{\alpha \beta}(Y,\tau^{\prime})  G_{0 i}(x^{\prime},\tf) \right\rangle \right|_{X^{\prime}=0, \mathbf{x}=0} \nonumber
    \\
    =& \frac{a^{4}\eb^{4}}{2} \int\limits_{K}  e^{i k_{0} \tau} \frac{e^{- 2\tf \tilde{K}^{2}}}{\tilde{K}^{2}}  \frac{\tilde{K}^{2}}{2}  \left[2 \tilde{k}_{0}^{2} + \tilde{K}^{2} \right]  \underbrace{\int\limits_{Q} \frac{\left(e^{- 2\tf \tilde{Q}^{2}}-1\right)}{\tilde{Q}^{2}}}_{\equiv \mathcal{I}_{3}(\tf)} \nonumber
    \\
     =& \frac{a^{4}\eb^{4}}{4} \int\limits_{K}  e^{i k_{0} \tau} e^{- 2\tf \tilde{K}^{2}}\left[2 \tilde{k}_{0}^{2} + \tilde{K}^{2} \right]   \mathcal{I}_{3}(\tf)\,,
\end{align}
where $L_{\beta}(Y,\tau^{\prime})$ is the Lagrange multiplier field needed to introduce the Wilson flow equation in its Lagrangian form, see \cite{Luscher:2011bx} for more details.
We note that in this case the integral of the leading-order contribution is not recovered, instead an additional power of $\tilde{K}^{2}$ emerges. Since this diagram represents a pure Wilson flow effect, it is not surprising that $\mathcal{I}_{3}(\tf)$ vanishes for zero flow time. The integral as a function of the flow time is shown in \autoref{fig:EE_LPT_QED}.
From the figure we can learn that the contribution of this diagram asymptotically reaches a constant value for large flow times. This is essentially a shift in the flow time that one can absorb into the definition of the flow time. To see this, we consider the case that more flow loops are attached to the propagator connecting the electric fields. These additional loops will lead to a correction of the form 
\begin{align}
    \text{LO}(\tau,\tf)\left( 1 + c_{1}(\tf) a^{2} \tilde{K}^{2} \phantom{\frac{ c_{2}(\tf) a^{4}}{2}} \right.&
    \\
   & \hspace{-10.0em} \left.+ \frac{ c_{2}(\tf) a^{4}}{2} \tilde{K}^{4} + \frac{ c_{3}(\tf) a^{6}}{6} \tilde{K}^{6} + \dots \right)\,, \nonumber
\end{align}
where the coefficient functions $c_{n}(\tf)$ are determined by the correlator at a given order in the lattice spacing $a$.  The sum of all diagrams with iterated
simple tadpoles of the sort shown in \autoref{fig:EE_LPT_QED} lead to $c_n(\tf)=(c_1(\tf))^n$,
and the corrections resum into the form
$e^{+ \tilde{K}^{2}  f(\tf)  }$, which can be interpreted as a finite order-$a^2$ renormalization of the flow time.
This effect has been observed before in previous lattice studies in~\cite{Cheng:2014jba,Hasenfratz:2014rna}.

Combining the contributions of the diagrams, we can now write down the full correlation function at next-to-leading order: 
\begin{align}
        G^{\textrm{\tiny NLO}}_{\textrm{\tiny QED}}(\tau,\tf)=&  \int\limits_{K} e^{i k_{0} \tau} \frac{e^{- 2\tf \tilde{K}^{2}}}{\tilde{K}^{2}} \left[ 2 \tilde{k}_{0}^{2}+ \tilde{K}^{2} \right]
          \\
        &\hspace{-3.5em} \times \left[ \eb^{2} + a^{4}\eb^{4} \left( -\frac{\mathcal{I}_{1} }{4} -\frac{3}{2} \mathcal{I}_{2}(\tf) +\frac{1}{4} \mathcal{I}_{3}(\tf) \tilde{K}^{2}      \right)  \right]\,. \nonumber
\end{align}
Evaluating this expression leads to an initial rise of the QED correlator as a function of flow time due to the renormalization of the electric field. We consider this effect to be present in the QCD color-electric correlator (\autoref{fig:EE_Nt16_Wilson}) as well, originating from similar renormalization effects as found in the NLO calculation in QED.

\end{document}